\documentclass[aps,prl,twocolumn,superscriptaddress]{revtex4-2}
\usepackage{amsmath, amsthm, amssymb}
\usepackage{mathrsfs}
\usepackage{graphicx}
\usepackage{dcolumn}
\usepackage{hyperref}
\usepackage{placeins}
\usepackage{amsfonts}
\usepackage{textcomp}
\usepackage{epstopdf}
\usepackage{hyperref}
\usepackage{color}
\usepackage{commath}
\usepackage{url}
\usepackage{gensymb}
\usepackage{siunitx}
\allowdisplaybreaks

\begin{document}
\title{Spontaneous generation of persistent activity in diffusively coupled cellular
assemblies}

\author{Ria~Ghosh}
\affiliation{The Institute of Mathematical Sciences, CIT Campus, 
Taramani, Chennai 600113, India}
\affiliation{Homi Bhabha National Institute, Anushaktinagar, Mumbai 
400 094, India}

\author{Shakti~N.~Menon}
\affiliation{The Institute of Mathematical Sciences, CIT Campus, 
Taramani, Chennai 600113, India}

\date{\today}

%----------------------------------------------------------------------------------------------%
\begin{abstract}
The spontaneous generation of electrical activity underpins a number of
essential physiological processes, and is observed even in tissues where
specialized pacemaker cells have not been identified. The emergence
of periodic oscillations in diffusively coupled assemblies of excitable and
electrically passive cells (which are individually incapable of sustaining
autonomous activity) has been suggested as a possible mechanism
underlying such phenomena. In this paper we investigate the dynamics
of such assemblies in more detail by considering simple motifs of coupled
electrically active and passive cells. The resulting behavior encompasses
a wide range of dynamical phenomena, including chaos. However,
embedding such assemblies in a lattice yields spatio-temporal patterns
that either correspond to a quiescent state or partial/globally synchronized
oscillations. The resulting reduction in dynamical complexity suggests an
emergent simplicity in the collective dynamics of such large, spatially
extended systems. Furthermore, we show that such patterns can be
reproduced by a reduced model comprising only excitatory and oscillatory
elements. Our results suggest a generalization of the mechanism by which
periodic activity can emerge in a heterogeneous system comprising
non-oscillatory elements by coupling them diffusively, provided their steady
states in isolation are sufficiently dissimilar.
\end{abstract}
%----------------------------------------------------------------------------------------------%

\maketitle

Spontaneously recurring electrical activity is of crucial significance in a number
of physiological contexts~\cite{Glass2001,Rabinovich2006,Sinha2014}. This is
typically driven by pacemaker cells~\cite{Huizinga1995,Thomson1998,Huizinga2014},
such as the sinoatrial node in the heart which comprises specialized cells that 
periodically generate signals initiating excitatory activity, leading to mechanical
contraction~\cite{Boyett2000}. However, such cells have not been observed in
other contractile tissue, such as the myometrium of the gravid
uterus~\cite{Smith2015}. It has been hypothesized that spontaneous activity in
the latter contexts arise through interactions between electrically active and
passive cells, local assemblies of which are capable of generating periodic
waves of activation in the tissue through diffusive
coupling~\cite{Cartwright2000,Boschi2001}. These waves traveling through an
organ are capable of sustaining spatio-temporally coherent
contractions~\cite{Singh2012}. Indeed it has been demonstrated that one of the
simplest ways to achieve this is by having an excitable cell coupled by
gap-junctions to one (or more) electrically passive cells characterized by a
resting state membrane potential that is much higher than that of the excitable
cell~\cite{Jacquemet2006}. The coupling between these heterogeneous cell
types causes the membrane potential of the excitable cell to be driven beyond
its threshold, resulting in the generation of an action potential. Subsequently,
the excitable cell attempts to return to its resting state, but after a period of
recovery is again driven to exceed its threshold by the passive cells coupled to
it, thereby resulting in a periodically recurring series of action potentials. Thus,
although neither excitable nor passive cells are individually capable of
spontaneous sustained activation, an assembly of these two cell types can
generate periodic oscillations~\cite{Quinn2016}. 

The emergence of periodic activity in a heterogeneous assembly of excitable
and passive cells makes such a mechanism a viable candidate for
self-organized system-wide coherent oscillations in physiological contexts
where no pacemakers have been reported~\cite{Wray2001,Young2018}.
Indeed, it has been demonstrated that a lattice of excitable cells that are each
coupled to a variable number of passive cells, can exhibit a range of
spatio-temporal phenomena consistent with those observed in the
uterus during the transition to coherent activity seen prior to
parturition~\cite{Singh2012,Xu2013,Xu2015}. However, noting that each
local cellular assembly are either in an excitable or an oscillatory dynamical
regime in isolation, raises an important question: 
can the observed collective behavior be reproduced in an even simpler
setting, viz., where each lattice site is occupied by either an oscillatory 
or an excitable element, a situation reminiscent of percolation~\cite{Stauffer1994}.
In this paper, we consider the dynamics of two classes of systems, each
capable of exhibiting spontaneous collective dynamics, one comprising 
electrically active and passive cells (EP) and the other comprising 
oscillatory and excitable cells (OE). We observe that simple motifs of cells
described using the EP model are capable of exhibiting a wide range of
complex collective dynamical patterns. However, several of these are no
longer observed when such cells are embedded in a spatially extended
system, suggesting an emergent simplicity of the collective dynamics. More
importantly, we observe that when cells described by the OE model are
placed on a lattice, the resulting dynamics are qualitatively very similar to
that obtained using the EP model. This points towards a more fundamental
mechanism that could explain the emergence of spontaneous recurrent
activity in physiologically relevant contexts.

To investigate in detail the range of complex behavior that emerges upon 
diffusively coupling excitable cells, each of which are in turn coupled to one
or more passive cells, we consider the simplest possible assemblies of
these cells capable of exhibiting spontaneous oscillatory activity. Following
Ref.~\cite{Singh2012}, we simulate the electrical activity of an excitable cell
using the FitzHugh-Nagumo (FHN) model~\cite{Keener1998}, which is
capable of both excitable and oscillatory dynamics. The model
describes the temporal evolution of an activation variable $V_e$ (the
membrane potential), and an inactivation variable $g$ (an effective
trans-membrane conductance) as $\dot{V}_{e} = \mathcal{F}(V_{e},g)$,
$\dot{g} = \mathcal{G}(V_{e},g)$. Here,
$\mathcal{F}(V_{e},g) = A\,V_{e}\,(V_e -\alpha)(1-V_e) - g$ and
$\mathcal{G}(V_{e},g) = \epsilon( k_{e}V_{e} - g - b )$, where $A(=3)$
and $k_e(=1)$ govern the kinetics, $\alpha(=0.2)$ is the excitation
threshold and $\epsilon(=0.08)$ is the recovery rate, while $b$
provides a measure of the asymmetry of the limit cycle. We note that
for $b_{c1} (=0.127) < b < b_{c2} (=0.343)$ the cell exhibits periodic
oscillations, while outside this range it is excitable with a stable resting
state. The value of $V_e$ in the resting state is close to $0$ for
$b<b_{c1}$ and consequently this regime is characterized as a ``low''
stable state. For $b>b_{c2}$, the resting state value of $V_e$ is
relatively large, with the regime being referred to as a ``high'' stable state.
The temporal evolution of the passive cell is described in terms of its
membrane potential $V_{p}$ as $\dot{V}_{p}= K\,(V_{p}^{R} - V_{p})$,
where $V_{p}^{R} (=1.5)$ is the resting state and $K (=0.25)$ is the
relaxation rate~\cite{Kohl1994}. Each excitable cell is electrically coupled
to $n_{p}$ ($= 0, 1, 2\ldots$) passive cells, where the conductance of the
gap junctions between the two cell types is $C_r$. Thus, the set of
equations used to describe the dynamics of an excitable cell $i$ coupled
to $n_{p}^i$ passive cells, as well as to an excitable cell $j$, is:
%----------------------------------------------------------------------------------------------%
\begin{equation}
\begin{split}
\frac{d V_{e}^i}{d t} &= \mathcal{F}(V_{e}^i,g^i) + n_{p}^i C_{r}(V_{p}^i-V_{e}^i) + D (V_{e}^j-V_{e}^i) \,,  \\
\frac{d g^i}{d t}     &= \mathcal{G}(V_{e}^i,g^i)\,, \\ 
\frac{d V_{p}^i}{d t} &= K(V_{p}^{R} -V_{p}^i)-C_{r}(V_{p}^i -V_{e}^i) \,.
\end{split}
\label{eq:fhn_passive}
\end{equation}
%----------------------------------------------------------------------------------------------%
The dynamics of an isolated excitable cell coupled to one or more passive
cells depends on $n_p$ and $C_r$. As $V_p^R$ is much higher than the
excitation threshold $\alpha$, for a range of $n_p$ and $C_r$ the coupled
excitable-passive system can exhibit oscillations. To demonstrate that such
emergent oscillations result exclusively from the coupling we have chosen
$b=0$, so that in isolation the FHN dynamics converges to the low stable
state. It is important to note in the context of the results reported here that
when $C_{r}>0.5$, oscillations are seen only for $n_p=1$ while those
excitable cells coupled to $n_p >1$ passive cells converge to high stable
states.
%----------------------------------------------------------------------------------------------%
\begin{figure}[tbp]
\begin{center}
\includegraphics[width=0.99\columnwidth]{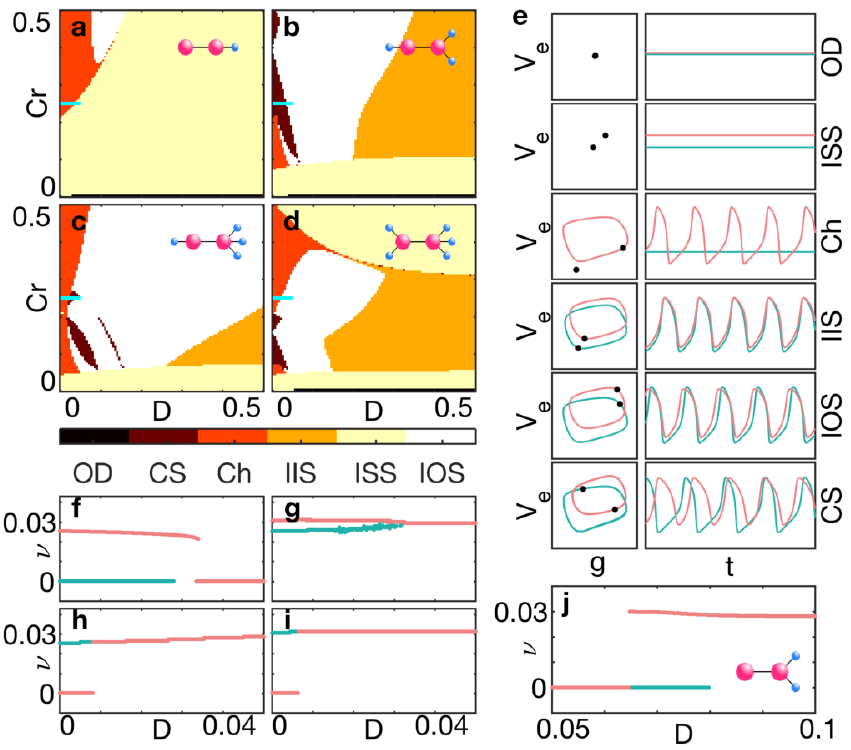}
\end{center}
\caption{\textbf{Emergent dynamics obtained with different motifs of
diffusively coupled excitable and passive cells.}
(a-d) Collective dynamical patterns observed over a range of values of
diffusive coupling strengths between two excitable cells ($D$) and
between an excitable and a passive cell ($C_r$), obtained using different
motifs (shown as insets in the corresponding panels, where the larger and
smaller circles represent excitable and passive cells, respectively). The
regimes are classified according to the dominant attractor obtained
for the given parameter set, and include oscillation death (OD),
inhomogeneous steady state (ISS), chimera (Ch),
inhomogeneous in-phase synchronization (IIS), inhomogeneous
out-of-phase synchronization (IOS) and cluster synchronization (CS).
(e) Phase plane trajectories and time series of the
excitable cells for the different collective dynamical regimes displayed in
(a-d). The black dots represent the instantaneous position of the two cells
on the corresponding limit cycle.
(f-i)~Variation of the frequency $\nu$ of the excitable cells on the
left (green) and right (maroon) in each of the
four motifs in (a-d) for $C_r=0.25$, over the range of $D$
indicated by horizontal cyan bars in (a-d).
As $D$ increases, frequencies of the two cells merge and for
sufficiently strong 
coupling the system can either stop oscillating (f), or display a frequency
that is between (g), greater than (h) or equal to (i) the maximum of the
intrinsic frequencies. (j) Variation of $\nu$ with $D$ for
$C_r=0.6$, obtained using the motif shown as an inset. Although both cells
are quiescent for low coupling strength, they exhibit oscillations for
sufficiently large $D$.
}
\label{fig_2units}
\end{figure}
%----------------------------------------------------------------------------------------------%
We first consider the simplest non-trivial assembly of dissimilar
excitable-passive units, viz., a pair of excitable cells diffusively
coupled with strength $D$, each interacting with a different number $n_p$
of passive cells with strength $C_r$ (Fig.~\ref{fig_2units}). The
heterogeneity in $n_p$ implies that the intrinsic behavior of the two units
are dissimilar, and we observe a range of distinct collective dynamics upon
varying $D$ and $C_{r}$, as shown in Fig.~\ref{fig_2units}~(a-d) for four
distinct connection topologies of the assemblies [illustrated in the top right
corners of the corresponding panels]. The dynamical regimes
obtained can be classified on the basis of the $V_e$ time-series of the
excitable cells, using a set of order parameters with specified threshold
values (see SI for details):
(i)~oscillation death (OD), where both cells are in the same temporally
invariant non-zero steady state;
(ii)~inhomogeneous steady state (ISS), where both cells are in different
temporally invariant steady states;
(iii)~chimera (Ch), where only one of the two cells oscillate; 
(iv)~inhomogeneous in-phase synchronization (IIS), where both cells
oscillate in-phase;
(v)~inhomogeneous out-of-phase synchronization (IOS), where both cells
have the same frequency but are out-of-phase with each other, and,
(vi)~cluster synchronization (CS), where the two cells have different
oscillation frequencies [Fig.~\ref{fig_2units}~(e)].
At low values of $D$, the two units can behave very
differently, and we observe collective states such as Ch or CS. As $D$
increases, the cells either become frequency locked or cease oscillating
altogether. Note that the intrinsic heterogeneity of the two units prevents
exact synchronization between them even for large $D$. For a given
value of $D$, as $C_r$ is decreased, eventually the cells stop oscillating
(in isolation, neither an excitable nor a passive cell is capable of spontaneous
activity).

In Fig.~\ref{fig_2units}~(f-i), we display the variation of the frequency
$\nu$ of the periodic activity exhibited by the excitable cells in the low
$D$ regime in each of the different assemblies ($C_r$ is fixed at $0.25$
in each case). We observe that an increase in $D$ either gives rise to the
cessation of oscillations [Fig.~\ref{fig_2units}(f)] or a synchronized state
in which the two units oscillate at a common frequency that is either lower
[Fig.~\ref{fig_2units}(g)], higher [Fig.~\ref{fig_2units}(h)] or equal to
[Fig.~\ref{fig_2units}(i)] the higher of the pair of intrinsic frequencies (i.e.,
the frequencies of each unit at $D=0$). Just as coupling an
excitable cell to one or more passive cells can, under appropriate
conditions, give rise to oscillatory dynamics, we observe that
spontaneous activity can arise upon coupling a pair of dissimilar units that
do not oscillate in isolation. Fig.~\ref{fig_2units}~(j)
shows a pair of excitable cells, having $n_p=0$ and $n_p=2$ respectively,
such that neither can independently oscillate for $C_r=0.6$.
However, upon increasing $D$ sufficiently, we observe a transition to
a Ch state and eventually to a frequency synchronized state of the two
units.

%----------------------------------------------------------------------------------------------%
\begin{figure}[tbp]
\begin{center}
\includegraphics[width=0.99\columnwidth]{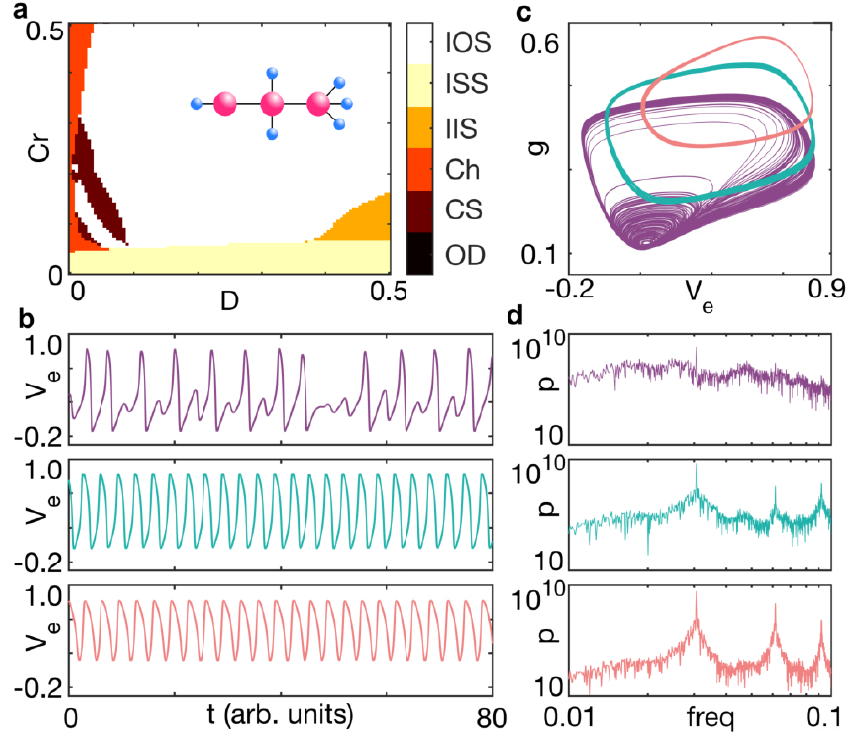}
\end{center}
\caption{
\textbf{Coexistence of chaotic and non-chaotic dynamical activity in a
system of three diffusively coupled excitable cells, each coupled to a
different number ($n_p$) of passive cells.}
(a) Collective dynamical patterns observed over a range of values of
diffusive coupling strengths $D$ between excitable cells and the
coupling $C_r$ between excitable and passive cells for the system
shown as an inset. The dynamical regimes are classified according to the
dominant attractor obtained for the given parameter set, and are the same
as those detailed in Fig.~\ref{fig_2units}~(a-d). (b) Time series of
membrane potential $V_e$ of the excitable cells coupled to (top)
$n_p=1$, (middle) $n_p=2$ and (bottom) $n_p=3$ passive cells, for
$D=0.02$ and $C_r=0.19$. (c-d) Phase plane trajectories and power
spectral densities of the three excitable cells, colored as in the
corresponding panels of (b). 
}
\label{fig_chaos}
\end{figure}
%----------------------------------------------------------------------------------------------%

Upon increasing the number of units in an assembly, we observe that the
system becomes capable of exhibiting more complex collective behavior
including chaotic activity. However, a particularly intriguing collective state
of coexisting chaotic and non-chaotic activity is observed in an assembly
of three excitable cells, having $n_p=1,2,3$ respectively, that are coupled
in a chain [see top right corner of Fig.~\ref{fig_chaos}~(a)]. For a large
range of values of $C_r$ and $D$, the system exhibits IOS
[Fig.~\ref{fig_chaos}~(a)]. However, in the CS regime,
we observe a collective dynamical state that is characterized by chaotic
behavior in the excitable cell with $n_p=1$ with non-chaotic, periodic
oscillations in the other two cells [Fig.~\ref{fig_chaos}~(b-d)]
The qualitative difference in the dynamics of the three excitable cells is
evident upon comparing their time series [Fig.~\ref{fig_chaos}(b)], phase
plane portraits [Fig.~\ref{fig_chaos}(c)] and power spectral densities
[Fig.~\ref{fig_chaos}(d)]. A more rigorous comparison, considering the
response of each cell to small perturbations, shows rapid divergence of
the resulting trajectory from the unperturbed one for the chaotic unit, with
no such deviation observed for the other two units (see SI). We note that
permutations of the connection topology of this assembly, i.e. changing the
order in which the cells with different values of $n_p$ are placed in the
chain, yields similar qualitative behavior, with chaotic dynamics
consistently observed in the unit with the lowest $n_p$.

\begin{figure}[tbp]
\begin{center}
\includegraphics[width=0.99\columnwidth]{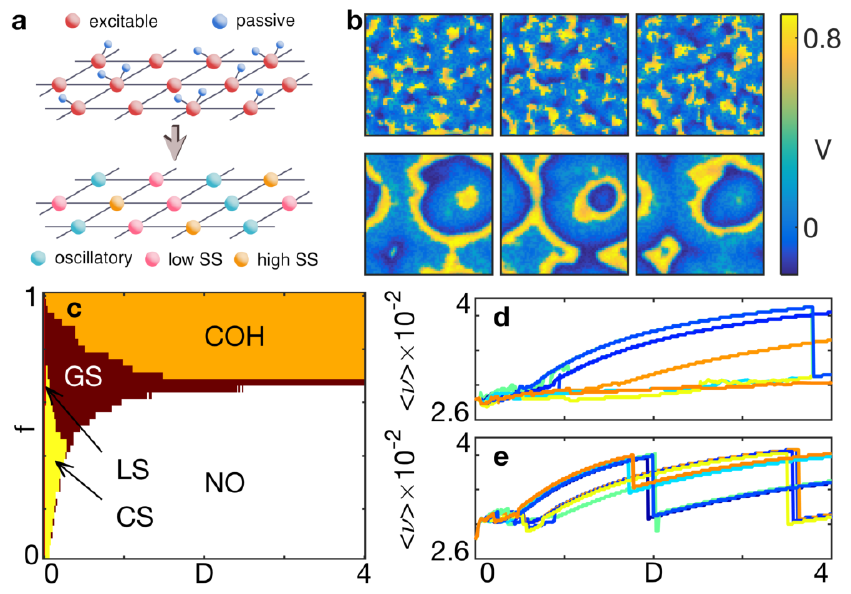}
\end{center}
\caption{
\textbf{Collective dynamics of a lattice of diffusively coupled elements
that can be either excitable or oscillatory.}
(a) Schematic diagram of uterine tissue, modeled as a two-dimensional
lattice where each site comprises an excitable cell coupled to a variable
number of passive cells (top, the ``EP'' model). The dynamics at each
site can equivalently be described through cells that are either oscillatory
or excitable (bottom, the ``OE'' model). The latter cell type could be
in one of two possible steady states, characterized by low and high
values of the state variable $V$. Note that the number of passive cells
coupled to each excitable cell in the top panel is chosen such that the
uncoupled dynamics at that site is qualitatively the same as that of the
corresponding site in the bottom panel.
(b) Snapshots of the activity $V$ in a planar simulation domain
comprising an equal mixture of excitable and oscillatory elements
($f=0.5$) diffusively coupled with strength $D$ to nearest neighbors,
showing (top row, $D=0.1$) cluster synchronization (CS) and (bottom
row, $D=0.3$) global synchronization (GS).
(c) Varying the diffusion coefficient $D$ and the fraction of oscillatory
cells $f$ in the lattice for the case of the OE model, several
distinct dynamical regimes are observed. In addition to CS and GS,
these include local synchronization (LS), coherence (COH) and no
oscillations (NO). The regions are identified according to the collective
dynamics  observed for the majority ($>50\%$) of initial conditions.
(d-e) Variation of the mean oscillatory frequency  $\langle\nu\rangle$ with
$D$. Each curve is obtained by starting from different random initial states
at low $D$ and then gradually increasing $D$ over time. In panel (d), the
cell at each site can be either oscillatory (with probability $f=0.7$) or
excitable (with probability $1-f$).  In panel (e), we first associate with
each lattice site a random number $n$ drawn from a
Poisson distribution (with mean $\lambda = 0.7$), and then place
oscillatory cells in sites with $n=1$ and excitable cells having a low or
high $V$ steady states at sites with $n=0$ and $n>1$, respectively. All
simulations are performed on square lattices comprising $64 \times 64$
cells, with periodic boundary conditions.
}
\label{fig_OE}
\end{figure}

It may appear that increasing the size of the assemblies further, by adding
more coupled excitable-passive units, can only lead to a further increase in the
complexity of the collective dynamics. However, surprisingly, we observe an
{\em emergent simplicity} in the behavior of large lattices of such units, with
neighboring elements coupled diffusively to each other. Indeed, such an
example is provided by a spatially extended model of uterine tissue, which
is heterogeneous by nature, comprising electrically active myocytes that
are excitable (thereby facilitating muscle contractions), as well as electrically
passive cells such as  interstitial Cajal-like cells (ICLC)~\cite{Duquette2005}
and fibroblasts [see top panel of Fig.~\ref{fig_OE}~(a)]. The system
exhibits spontaneous oscillations for a range of values of $C_r$ even though,
in isolation, none of the individual cells are capable of autonomous periodic
activity, as has been experimentally observed in uterine
tissue~\cite{Wray2001,Young2018}. More important from the perspective of
the dynamical transition to periodic coordinated contraction of the myometrium,
it is seen that increasing $D$ results in the self-organized emergence of
global synchronization, and eventually coherence~\cite{Singh2012,Xu2015}.
It is striking that such coordination is achieved exclusively through local
interactions between cells and does not require a centralized 
pacemaker such as that present in the heart (viz., the sino-atrial node).

%----------------------------------------------------------------------------------------------%
%----------------------------------------------------------------------------------------------%

The relative simplicity of the collective behavior of such a lattice of
heterogeneous coupled cells can be shown by demonstrating that it can be
captured by a reduced description of the system in terms of interacting
dynamical elements, each of which are either in an oscillating or a steady state.
In particular, we can replace excitable-passive cell assemblies that are capable
of spontaneous periodic activation by a single FHN unit with $b_{c1} < b < b_{c2}$
(for concreteness, we choose $b_{osc} =0.192$ for the simulations whose results
are shown here), and FHN units with $b<b_{c1}$ ($>b_{c2}$) for cell assemblies
exhibiting a low (high) stable state (we choose $b_{exc}^{low}=0$ and
$b_{exc}^{high}=0.394$ for the simulations shown here). The resulting equivalent
lattice now comprises only FHN units, a fraction $f$ of which are in an oscillatory
regime with the remaining being excitable by virtue of having different values of
$b$ [see Fig.~\ref{fig_OE}~(a), bottom panel]. Nevertheless,
the system exhibits qualitatively identical behavior to that seen in models of
uterine tissue simulated by coupling assemblies of excitable and passive
elements, e.g., the occurrence of cluster synchronization at relatively low
inter-cellular coupling that gives way to global synchronization of periodic activity
(coordinated by propagating waves of excitation that traverse the lattice) for
stronger coupling [Fig.~\ref{fig_OE}~(b)]. 

The similarity of the emergent properties of the simpler model can be established
further by comparing the different dynamical regimes of the $f-D$ parameter space
with that observed in the uterine model having heterogeneous cell
types~\cite{Singh2012}. Indeed all the qualitatively distinct types of behavior
reported in the latter can be seen in Fig.~\ref{fig_OE}~(c), including No Oscillation
(NO, with all cells in steady states), Cluster Synchronization (CS, marked by
coexistence of multiple groups of cells, each characterized by a different
frequency), Local Synchronization (LS, coexistence of quiescent cells with cells
oscillating at a common frequency), Global Synchronization (GS, all cells
have the same oscillation frequency) and Coherence (COH, all cells exhibit
phase synchrony). In the limit of large $D$, the dynamics of lattice can be further
simplified and an implicit analytical equation can be obtained for $f_c$, the fraction of
FHN units that should be oscillatory for the system to exhibit persistent periodic
excitations. It marks the boundary between the NO and COH regimes and is
given as $b_{c1} =(1-f_{c})\,b_{exc} + f_{c}\,b_{osc}$. For the
situation shown in Fig.~\ref{fig_OE}~(c), $b_{exc}=b_{exc}^{low}$, which yields
$f_c \sim 0.7$ upon inserting the corresponding numerical values.

We can investigate the collective dynamics around this asymptotic
boundary between persistent oscillatory activity and a quiescent steady
state by considering the case $f=0.7$. In particular, we focus on the variation of
the overall activation rate, measured by the mean frequency of periodic activation
$\langle\nu\rangle$ (averaged over all oscillating elements in the lattice), as $D$
is increased. In order to be consistent with the physiological setting, where the 
coupling between cells increases over the gestation period (as a result of hormone 
induced increased expression of gap junctions that electrically couple the 
cells~\cite{Jahn1995}), $D$ is increased adiabatically over the course of a
single realization, from a random initial condition at a low value of $D$. For the
situation when $b_{exc}=b_{exc}^{low}$, shown in Fig.~\ref{fig_OE}~(d),
$\langle\nu\rangle$ increases with $D$ until it reaches a maximum value related
to the reciprocal of the refractory period (set by the parameters of the FHN
model). Increasing $D$ further results in an abrupt drop in $\langle\nu\rangle$ as 
the number of propagating wavefronts in the system changes. A subsequent
increase in $D$ results in an increase in $\langle\nu\rangle$
generated by the new spatio-temporal pattern. This is qualitatively the same as
the phenomenon observed for the model of uterine tissue involving
assemblies of excitable and passive cells. An even closer match between
the two classes of models can be obtained if we replace each of the excitable
elements with FHN elements having either $b_{exc}=b_{exc}^{low}$ or
$=b_{exc}^{high}$ according to the following procedure: first, each lattice site
is assigned a value $n$ chosen from a Poisson distribution with mean $\lambda=f$.
Note that this is identical to the process by which the number of passive cells 
(given by $n$) coupled to an excitable cell are determined in modeling uterine
tissue with excitable-passive cell assemblies~\cite{Singh2012}. Next, FHN
units in the oscillatory regime ($b=b_{osc}$) are placed at sites having $n=1$,
while FHN units with $b=b_{exc}^{low}$ (i.e., excitable element with a low
stable state) are placed at sites with $n=0$. At sites having $n>1$, corresponding
to excitable-passive cell assemblies whose activity is arrested at a high stable
state, FHN units with $b=b_{exc}^{high}$ are placed. The resulting
oscillatory-excitable (OE) model can accurately reproduce dynamical
behaviors reported for the model comprising excitable-passive (EP)
cell assemblies~\cite{Singh2012}. These include the emergence of clusters
characterized by a common oscillation frequency, propagating wavefronts,
as well as self-sustained spiral waves in the GS regime [see SI for details]. Note
that persistent periodic activity can arise upon coupling two FHN units that cannot 
independently oscillate, provided one of them is in the low and the other in the high 
stable state - a phenomenon analogous to the appearance of oscillations in
assemblies of excitable and passive cells which cannot sustain autonomous
activity. Fig.~\ref{fig_OE}~(e) shows the evolution of the mean frequency with
$D$ when the cellular coupling is increased adiabatically starting from a random
initial condition over the course of a single realization, displaying an even closer
agreement to the behavior seen in the EP model of uterine tissue~\cite{Singh2012}.

The qualitative equivalence of the collective behavior in large lattices for
the two classes of models is all the more surprising as the dynamics of
network motifs comprising excitable-passive cell assemblies (discussed
above, see Figs.~\ref{fig_2units} and~\ref{fig_chaos}) is much more complex
than that observed upon replacing each assembly by a FHN unit in the oscillatory
or excitable regime. For instance, coupling a pair of EP cell assemblies,
each of which oscillate at different frequencies, cannot give rise to
exact synchronization even at large $D$. However, two FHN
oscillators characterized by distinct $b$ values (and hence, frequencies) 
can exhibit exact synchronization when coupled with sufficient
strength. Furthermore, while we have reported motifs of connected
EP cell assemblies that exhibit chimera (Ch) regimes over a range of
coupling strengths, such behavior cannot be seen in two coupled FHN
oscillators with distinct intrinsic frequencies. We would also like to point out
that nothing equivalent to the chaotic behavior observed in a motif
comprising coupled EP cell assemblies (Fig.~\ref{fig_chaos}) is seen in
systems of coupled FHN oscillators arranged in a similar topology (viz., a
chain comprising two oscillators having different intrinsic frequencies and
an excitable element). Thus, even though the OE model reproduces the
collective behavior of a large system of coupled EP cell assemblies, the
dynamics at the microscopic level (i.e., motifs comprising only a few elements)
can be extremely different for the two classes of models (see SI).

To conclude, in this paper we have shown that while coupled excitable-passive cell
assemblies are capable of exhibiting a wide range of dynamical behaviors including
chaos, a macroscopic system comprising a large number of such elements
diffusively coupled to their nearest neighbors on a lattice shows relatively simpler
spatio-temporal phenomena. In particular, this resulting collective dynamics can be
reproduced by a model comprising many elements, each described by a generic
model for an excitable cell that is either in a steady state or in an oscillatory
regime. Indeed, it suggests that the behavior associated with physiologically detailed
models of uterine tissue activity~\cite{Tong2011,Xu2013,Xu2015} can be understood
in terms of a reduced model involving a heterogeneous assembly of coupled
oscillatory and excitable elements. More importantly, our results point towards a
generalization of the mechanism proposed in Ref.~\cite{Jacquemet2006} for the 
emergence of periodic activity in systems where none of the individual elements are
intrinsically capable of oscillating. While it was shown there that persistent
oscillations arise upon coupling excitable and electrically passive cells under
certain circumstances, here we have shown that an oscillating system may emerge
upon coupling elements, each of which are in isolation at time-invariant steady states
- provided these states are dissimilar (i.e., the state variables associated with them
have sufficiently distinct numerical values, corresponding to ``low'' and ``high''). 
Furthermore, our demonstration of a large variety of dynamical attractors in small
assemblies of excitable and passive elements can provide an understanding of the
complex dynamics seen in electrically coupled heterogeneous sub-cellular
compartments in neurons~\cite{Safronov2000,Bekkers2007} and small networks of
neurons interacting via gap-junctions~\cite{Gutierrez2013}.

\begin{acknowledgments}
We would like to thank Sitabhra Sinha and K.~A.~Chandrashekar for helpful 
discussions. SNM has been supported by the IMSc Complex Systems Project
(12th Plan), and the Center of Excellence in Complex Systems and Data 
Science, both funded by the Department of Atomic Energy, Government of 
India. The simulations and computations required for this work were
supported by the Institute of Mathematical Sciences High Performance
Computing facility (hpc.imsc.res.in) [Nandadevi and Satpura clusters]. 
\end{acknowledgments}

%-----------------------------------------------------------------------------------------------------------%

%-----------------------------------------------------------------------------------------------------------%

\clearpage

%=============================================================%
\onecolumngrid

\setcounter{figure}{0}
\renewcommand\thefigure{S\arabic{figure}}  
\renewcommand\thetable{S\arabic{table}}

\vspace{1cm}

\begin{center}
\large{
\uppercase{Supporting Information for}\\ 
\vspace{1em} 
Spontaneous generation of persistent activity in diffusively coupled cellular assemblies
}
\end{center}

\vspace{0.5cm}
\section*{List of Supplementary Figures}

%--------------------------------------------------------------------------------------------------------------%
\begin{enumerate}

\item Decision tree to identify the dynamical state of a system of coupled excitable 
cells described by the FitzHugh-Nagumo model, each of which are coupled to a 
variable number of passive cells (EP model).

\item Decision tree to identify the dynamical state of a heterogeneous system of 
coupled cells described by the FitzHugh-Nagumo model, some of which are in the 
oscillatory regime, with the remaining ones in the excitable regime (OE model).

\item Differential sensitivity to small perturbations in a system of three coupled
excitable cells, each attached to a variable number of passive cells, that displays
coexistence of chaotic and non-chaotic activity for $C_r=0.19$ and $D=0.02$, as 
described in the main text.

\item Time-evolution of the activation variables for two nodes coupled to each other in
the EP model and the OE model, showing the absence and presence, respectively, of 
exact synchronization, and the spontaneous generation of activity in the OE model. 

\item Comparison of spatio-temporal evolution of the activity in two-dimensional
systems of the EP model and the OE model.

\end{enumerate}
%--------------------------------------------------------------------------------------------------------------%

\FloatBarrier

\vspace{0.5cm}
\section*{Identifying the collective dynamical states of the models}

\begin{figure}[tbp]
\begin{center}
\includegraphics[width=0.99\columnwidth]{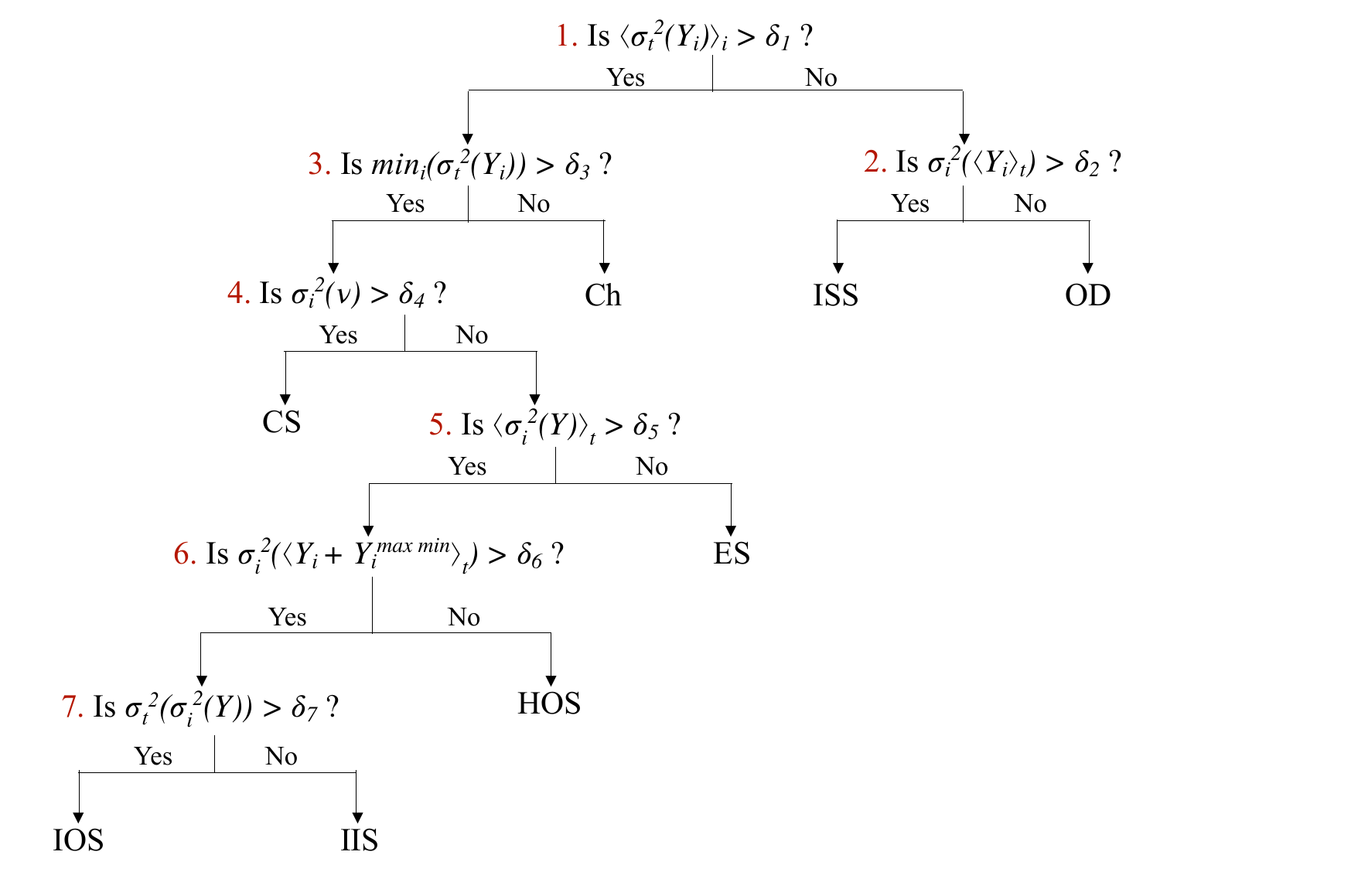}
\end{center}
\caption{Decision tree to identify the dynamical state of a system of coupled excitable
cells described by the FitzHugh-Nagumo model, each of which are coupled to a 
variable number of passive cells (EP model). The order parameters used for 
determining the nature of the spatio-temporal pattern observed include the 
dispersions calculated across space [$\sigma_i^2 (~)$], or across time
[$\sigma_t^2 (~)$], for the state variables $Y_i$ ($i=1,\ldots,N$) or the oscillation 
frequencies $\nu$. Averages calculated over space and over time are denoted by
$\langle~\rangle_i$ and $\langle~\rangle_t$ respectively, while the deviation of the 
lowest value of the state variable for any node from its maximum calculated across 
space is denoted by $Y_i^{\max\min}=\max_i(\min_t(Y_i))-\min_t(Y_i).$ The threshold 
values used to distinguish between the different states, viz., Oscillation Death (OD),
Inhomogeneous Steady State (ISS), Chimera (Ch), Cluster Synchronization (CS), 
Exact Synchronization (ES), Homogeneous Out-of-phase Synchronization (HOS), 
Inhomogeneous In-phase Synchronization (IIS), and Inhomogeneous Out-of-phase 
Synchronization (IOS), are:
$\delta_1=10^{-3}, \; \delta_2=10^{-5}, \; \delta_3=10^{-3}, \; \delta_4=10^{-7}, \; \delta_5=10^{-5}, \; \delta_6=10^{-7},$ and $\delta_7=10^{-4}$.
} 
\label{fig_OP_motif}
\end{figure}

\begin{figure}[tbp]
\begin{center}
\includegraphics[width=0.99\columnwidth]{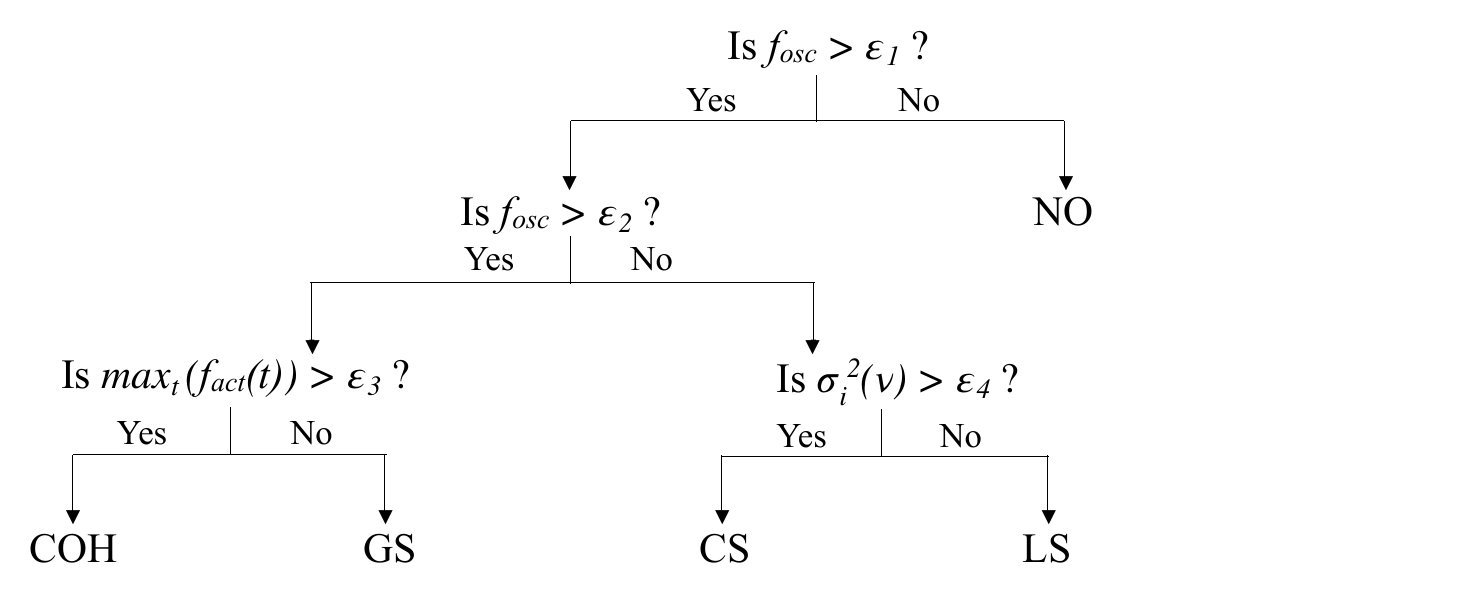}
\end{center}
\caption{Decision tree to identify the dynamical state of a heterogeneous system of 
coupled cells described by the FitzHugh-Nagumo model, some of which are in the 
oscillatory regime with the remaining ones in the excitable regime (OE model). The 
order parameters used for determining the nature of the spatio-temporal pattern 
observed include (i) $f_{osc}$: the number of oscillating nodes in the lattice,
(ii)~$max_t(f_{act}(t))$: the maximum number of nodes that were active (i.e., above
the excitation threshold $\alpha$) together over the period of observation, and
(iii)~$\sigma_i^2(\nu)$ is the dispersion of frequencies of all oscillating nodes 
calculated across space. The threshold values used to distinguish between the 
different states, viz., No Oscillations (NO), Cluster Synchronization (CS), Local 
Synchronization (LS), Global Synchronization (GS), and Coherence (COH), are:
$\epsilon_1=10^{-3}, \; \epsilon_2=0.999, \; \epsilon_3=0.995$ and
$\epsilon_4=10^{-10}$.
}
\label{fig_OP_percol}
\end{figure}

\noindent
To characterize the various spatiotemporal patterns of collective dynamics that are
observed in the models investigated in the main text, we measure several order 
parameters to aid in distinguishing the states. This is done with the help of the 
decision trees shown in Fig.~\ref{fig_OP_motif} (for the EP model) and 
Fig.~\ref{fig_OP_percol} (for the OE model). At each numbered decision point, 
threshold values $\delta_{1,\ldots,7}$ (Fig.~\ref{fig_OP_motif}) or
$\epsilon_{1,\ldots,4}$ (Fig.~\ref{fig_OP_percol}), whose numerical values are 
indicated in the respective captions, are used to determine the answer to the 
corresponding question.

%--------------------------------------------------------------------------------------------------------------%
%--------------------------------------------------------------------------------------------------------------%

\vspace{0.5cm}
For the EP model comprising $N$ coupled nodes (Fig.~\ref{fig_OP_motif}), where the 
order parameters are calculated using the time-series $Y_i (t)$ representing the 
activation variable $V_e$ of $i$-th node $(i=1,\ldots, N)$, the questions asked at the 
different decision points are:
%--------------------------------------------------------------------------------------------------------------%
\begin{enumerate}
\item{Is there temporal variation in $Y$~? In practice, we measure the dispersion of 
each time-series $Y_i$ and check if the average of this quantity over the nodes is 
greater than a threshold $\delta_1 \ll 1$. If true, at least one of the nodes is oscillating; 
else, the nodes are in time-invariant states.}

\item{If the nodes are in a time-invariant state, are the values of $Y_i$ identical for
all $i$~? In practice, we calculate the dispersion of the mean of the time-series across
the nodes and check if it exceeds a threshold $\delta_2 \sim 0$. If true, it indicates
that at least some of the $Y_i$ are different, which characterizes an Inhomogeneous 
Steady State (ISS); else, all $Y_i$ are the same indicating that it is an Oscillator 
Death (OD) state.}

\item{If there is temporal variation in $Y$, are all nodes oscillating~? In practice, we
check if the smallest dispersion of $Y$ calculated for all nodes is greater than a
threshold $\delta_3 \ll 1$. If not satisfied, it implies that at least one of the nodes is
not oscillating and hence the state corresponds to a Chimera (Ch).}

\item{If all nodes are oscillating, do they all the same frequency~? In practice, we 
check if the dispersion of the oscillation frequencies is greater than a threshold
$\delta_4 \sim 0$. If true, then it corresponds to nodes having distinct frequencies
which characterizes the Cluster Synchronization (CS) state.}

\item{If all nodes are oscillating at the same frequency, are they oscillating in phase~?
In practice, we measure the instantaneous dispersion between the amplitudes of the
different nodes and check if its temporal average exceeds the threshold
$\delta_5 \sim 0$. If it does not, the state corresponds to Exact Synchronization (ES) 
(belonging to the broader category of coherent states).}

\item{If the oscillators are not in phase, are their amplitudes the same~? For this, we
vertically displace each time series $Y_i$ such that their minima $min(Y_i)$ are the 
same. We then calculate the dispersion in the amplitude and check if the difference is 
greater than the threshold $\delta_6 \sim 0$. If not, it corresponds to Homogeneous 
Out of phase Synchronization (HOS), where the oscillations have the same amplitude 
but are not phase synchronized (one of the two states that belong to the broader 
class of Global Synchronization).}

\item{If the oscillators have different amplitudes, are they phase synchronized~? This 
can be determined by asking if the dispersion of $Y$ calculated over the different 
nodes is invariant in time. In practice we calculate the dispersion of this quantity over 
time and check if it is greater than the threshold $\delta_7 \ll 1$. If not satisfied, then 
the nodes are synchronized in phase although oscillating with different amplitudes,
corresponding to the state of Inhomogeneous In-phase Synchronization (IIS) 
(belonging to the broader category of coherent states); else, the state is
Inhomogeneous Out of phase Synchronization (IOS) (the other state that belongs to 
the broader class of Global Synchronization).}

\end{enumerate}
%--------------------------------------------------------------------------------------------------------------%
Note that we do not observe ES or HOS for the range of parameters used for the 
results reported in this paper.\\
\vspace{0.2cm}

\noindent
For the OE model (Fig.~\ref{fig_OP_percol}), the questions asked in the different 
decision points are:
%--------------------------------------------------------------------------------------------------------------%
\begin{enumerate}
\item Is there a finite number of oscillating nodes in the system~? In practice, we ask 
if the fraction of oscillating nodes $f_{osc}$ is above a threshold $\epsilon_1 \ll 1$. 
If this is not satisfied, we classify the state as No Oscillations (NO).

\item Are all nodes oscillating~? Again, in practice, we determine if $f_{osc}$ is 
greater than a threshold $\epsilon_2 \sim 1$. If the answer is yes, we go on to ask if 
the oscillations are synchronized, while if the answer is no, we ask if the frequencies 
of the oscillating nodes are identical.

\item Are all the oscillating nodes synchronized in their phase~? This is true if almost
the entire set of nodes were simultaneously active (a node is considered to be active 
if its activation variable $V_e$ is above the excitation threshold) and determined in
practice by checking if the fraction of active nodes is larger than a threshold
$\epsilon_3 \sim 1$. If true, then the states corresponds to Coherence (COH), else it 
is labeled as Global Synchronization (GS).

\item Are the frequencies of the oscillating nodes identical~? We measure the 
dispersion of the frequencies and check if this is greater than a threshold
$\epsilon_4 \sim 0$. If true, then the state is labeled as Cluster Synchronization (CS) 
characterized by the existence of many groups of oscillators having distinct 
frequencies; otherwise, the state is referred to as Local Synchronization (LS) in which 
all oscillating nodes have the same frequency.
\end{enumerate}
%--------------------------------------------------------------------------------------------------------------%

%--------------------------------------------------------------------------------------------------------------%
%--------------------------------------------------------------------------------------------------------------%

%==============================================================%
\section*{Characterizing the coexistence of chaotic and non-chaotic dynamics in a 
$3$-node motif of the EP model}

%--------------------------------------------------------------------------------------------------------------%
\begin{figure}[tbp]
\begin{center}
\includegraphics[width=0.99\columnwidth]{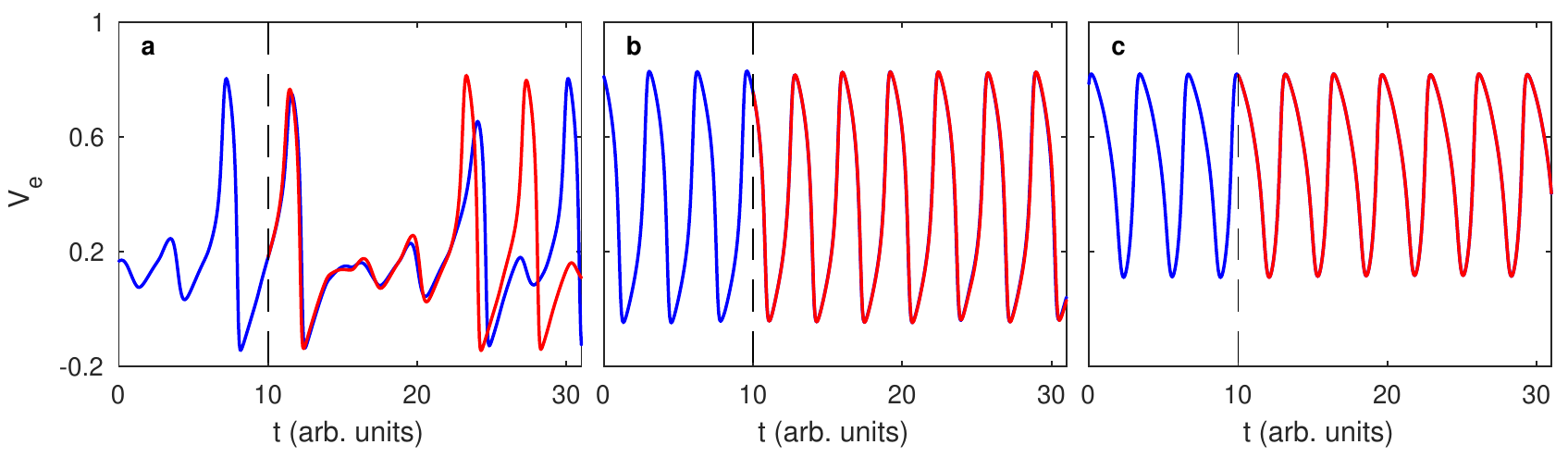}
\end{center}
\caption{Differential sensitivity to small perturbations in a system of three coupled 
excitable cells, each attached to a variable number of passive cells, which display 
coexistence of chaotic and non-chaotic activity for $C_r=0.19$ and $D=0.02$ as 
described in the main text [see Fig.~2]. Each cell is subjected to a perturbation of 
the same magnitude ($=10^{-3}$) applied at $t=10$ arb. units. While the cell 
exhibiting chaotic activity [panel (a)] displays exponential divergence of the perturbed 
trajectory (red) from the original one (blue), in the other two cells showing non-chaotic 
activity [panels (b) and (c)] the perturbed and original trajectories almost coincide.
}
\label{fig_SIchaos}
\end{figure}
%--------------------------------------------------------------------------------------------------------------%

In the main text, we report the coexistence of chaotic and non-chaotic activity in a
motif comprising $3$ FHN nodes coupled to $n_p = 1, 2$ and $3$ passive cells
respectively (see Fig.~2). We note that the usual method of demonstrating chaotic 
activity in a system is to show that its dynamics is sensitively dependent on initial 
conditions. In other words, two trajectories that begin from points that lie very close to 
each other in phase space will exponentially diverge with time. This is demonstrated 
in (Fig.~\ref{fig_SIchaos}) where the dynamical state of each unit is independently 
altered in turn by a very small magnitude perturbation and the subsequent time-
evolution (shown in red) is compared with the trajectory of the node in the 
unperturbed system (shown in blue). As can be seen in panel (a), in the node 
exhibiting chaotic activity, the perturbed trajectory rapidly deviates from that observed 
in absence of perturbation, while for the other two nodes which display periodic 
activity, the perturbation does not result in any perceptible deviation [panels (b) and 
(c)]. The chaotic nature of the dynamics can be quantitatively established by noting 
that the maximal Lyapunov exponent (calculated by the TISEAN software) is positive.

%==============================================================%
\section*{Comparing the collective dynamics observed in the OE and EP models}

%--------------------------------------------------------------------------------------------------------------%
\begin{figure}[tbp]
\begin{center}
\includegraphics[width=0.99\columnwidth]{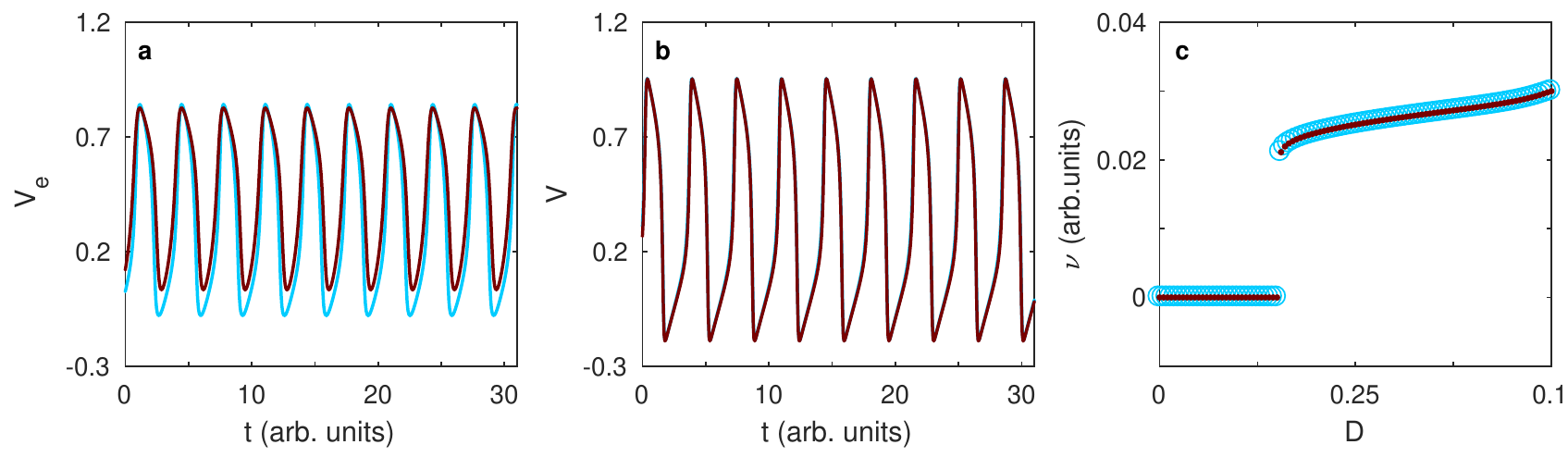}
\end{center}
\caption{Time-evolution of the activation variables for two nodes coupled to each
other in (a) the EP model and (b) the OE model, showing the absence and presence,
respectively, of exact synchronization. In (a), the two excitable cells are coupled to
$n_p=1$ (blue curve) and $n_p = 2$ (red curve) passive cells, respectively, with 
strength $C_r=0.4$ and interacting with each other with strength $D=0.2$. In (b) the 
two cells differ in the numerical value of the FHN model parameter $b$, which are 
$0.2$ and $0.181$ respectively. (c) Oscillation frequencies for a system of two 
coupled excitable cells, each described by the FHN model but with distinct values for 
the parameter $b$ ($=0$ and $0.394$, respectively) such that they have very different 
steady states in isolation. The system is quiescent in the absence of any external 
stimulation until the coupling strength exceeds a critical value, when spontaneous 
periodic activity is observed.
}
\label{fig_OE_EP_TS}
\end{figure}
%--------------------------------------------------------------------------------------------------------------%

\begin{figure}[tbp]
\begin{center}
\includegraphics[width=0.99\columnwidth]{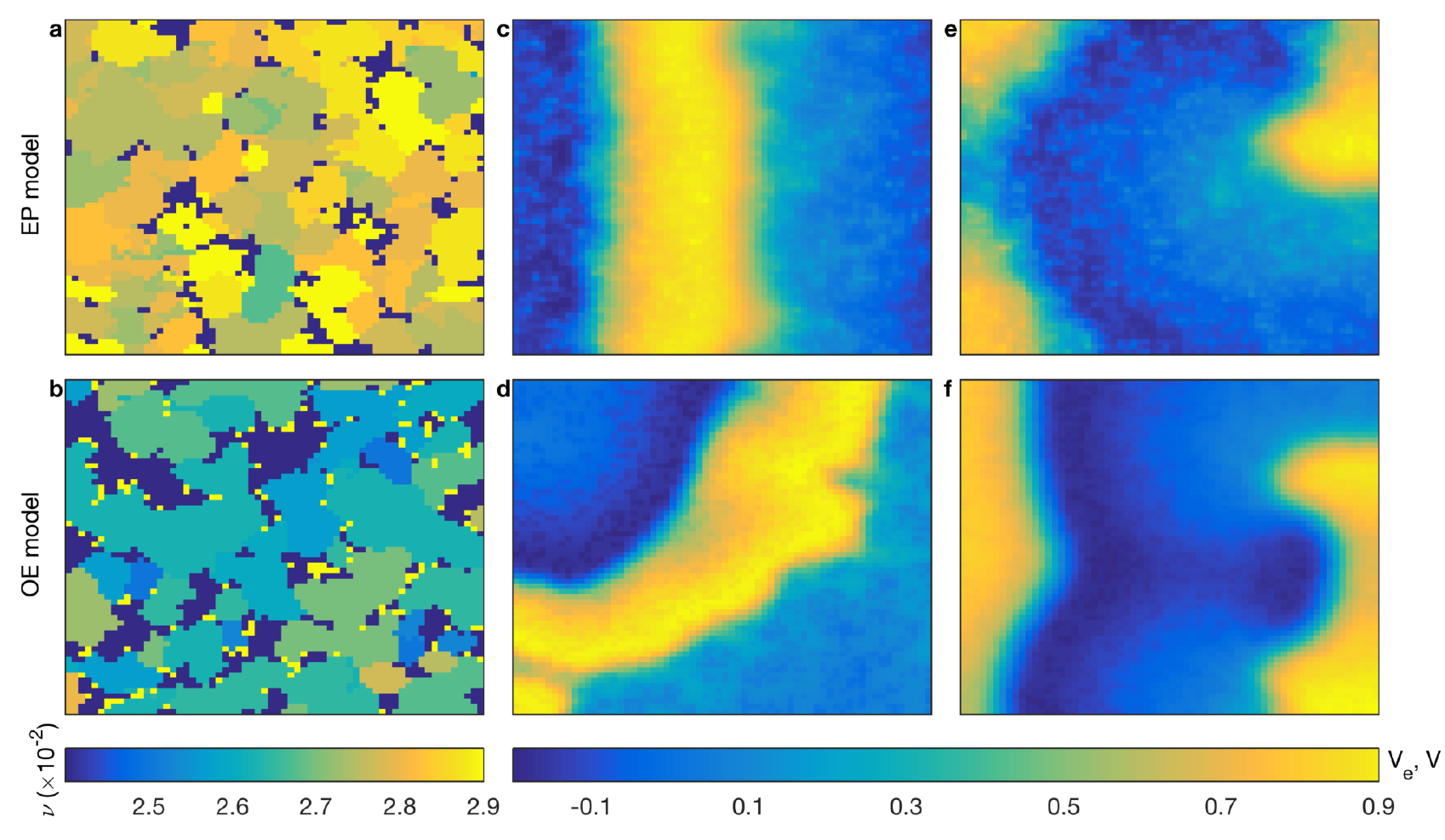}
\end{center}
\caption{Comparison of spatio-temporal evolution of the activity in
two-dimensional systems of the EP model (top row, $V_e$) and the OE model
(bottom row, $V$), showing similar behavior, viz., (a,b) cluster synchronization, as 
evident from the spatial distribution of frequencies $\nu$,
(c,d) travelling wavefronts, and (e,f) spiral waves.
}
\label{fig_OE_EP_compare}
\end{figure}

In this subsection, we show that similar collective dynamical behavior is displayed by 
both models (Figs.~\ref{fig_OE_EP_TS} and ~\ref{fig_OE_EP_compare}). As both 
models use the FHN model to describe the individual excitable nodes, its parameters 
$A$, $\alpha$, $\epsilon$ and $k_e$ are kept the same in the two models such that it 
is a fair comparison. However, values for the distinct parameters in the two models, 
viz., $b$ in the OE model and $C_r, n_p$ in the EP model, need to be suitably 
chosen such that a correspondence can be maintained between them. We note that 
for a given value of $C_r$, the resting state or the oscillation frequency of a node in 
the EP model depends on the value of $n_p$. On the other hand, in the OE model, it 
is the value of $b$ which determines these. Thus, we can relate values of $\{C_r,np\}$ 
with those of $b$ that give rise to the same resting state or oscillation frequencies in 
the two models, respectively. \\

%--------------------------------------------------------------------------------------------------------------%
%--------------------------------------------------------------------------------------------------------------%

Fig.~\ref{fig_OE_EP_TS}~(a) shows the time-series of the activation variable $V_e$ 
for a pair of coupled excitable units in the EP model, one connected to $n_p= 1$ 
passive cell (shown in blue) and the other to $n_p=2$ passive cells (shown in red). 
We observe that while the two are phase synchronized, their amplitudes are different. 
In the OE model shown in Fig.~\ref{fig_OE_EP_TS}~(b), if we couple two FHN cells 
having distinct $b$ values such that the coupled system has the same frequency as 
the EP model, we observe exact synchronization, as is evident from the complete 
overlap of the two time-series. Thus, not only are two nodes in the OE model phase 
synchronized (as is also the case in the EP model), they also have identical 
amplitudes (unlike in the EP model). Fig.~\ref{fig_OE_EP_TS}~(c) shows the variation 
of the oscillation frequency $\nu$ as a function of the coupling strength $D$ between 
two FHN nodes having $b=0$ and $0.394$, respectively, in the OE model. The 
uncoupled cells ($D=0$) do not show spontaneous oscillations - however, when they 
are coupled with a sufficient strength $D$ they exhibit synchronized oscillations. The 
emergence of spontaneous oscillatory activity in the coupled system is similar to that 
reported for the EP model motifs in the main text [e.g., see Fig.~1~(j)].\\

We also observe similar dynamical regimes in the two models when considering a 
large number of coupled units arranged in two-dimensional lattices 
(Fig.~\ref{fig_OE_EP_compare}). For instance, the collective dynamical states seen
in the EP model, e.g., cluster synchronization (characterized by several groups such 
that the nodes in each are oscillating at a common frequency that is distinct from that 
of other groups) are also observed in the OE model [compare panels (a) and (b)]. 
Similarly, travelling wavefronts and spiral waves  that are observed in the global 
synchronization regime in the EP model for annealed simulations [see panels (c)
an (e)] are also observed in the OE model under suitable conditions [see panels (d) 
and (f)].

\end{document}